\documentclass[paper]{elsarticle}

\usepackage{epsfig,color}

\begin{document}
\title{{\Large \textbf{Bosonic binary mixtures with Josephson-type
interactions}}}
\author{Val\'eria de C. Souza}
\address{Departamento de F{\'\i}sica Te\'orica, Instituto de F\'{\i}sica,
Universidade do Estado do Rio de Janeiro, Rua S\~ao Francisco Xavier 524, 20550-013,  Rio de Janeiro, RJ, Brazil.}
\author{Zochil Gonz\'alez Arenas}
\address{Departamento de Matem\'atica Aplicada,Instituto de Matem\'atica e Estat\'\i stica,  
Universidade do Estado do Rio de Janeiro, Rua S\~ao Francisco Xavier 524, 20550-900,  Rio de Janeiro, RJ, Brazil.}
\author{Daniel G.\ Barci and Cesar A. Linhares}
\address{Departamento de F{\'\i}sica Te\'orica, Instituto de F\'{\i}sica,
Universidade do Estado do Rio de Janeiro, Rua S\~ao Francisco Xavier 524, 20550-013,  Rio de Janeiro, RJ, Brazil.}

\date{\today}

\begin{abstract}
Motivated by experiments in bosonic mixtures composed of a single element in
two different hyperfine states, we study bosonic binary mixtures in the
presence of Josephson interactions between species. We focus on a particular
model with $O(2)$ isospin symmetry, lifted by an imbalanced population
parametrized by a Rabi frequency, $\Omega _{R}$, and a detuning, $\nu $,
which couples the phases of both species. We have studied the model at
mean-field approximation plus Gaussian fluctuations. We have found that both
species simultaneously condensate below a critical temperature $T_{c}$ and
the relative phases are locked by the applied laser  phase, $\alpha$.
Moreover, the condensate fractions are strongly dependent on the ratio 
$\Omega _{R}/|\nu |$ that is not affected by thermal fluctuations. 
\end{abstract}


\maketitle

\section{Introduction}
Multicomponent quantum gases are fascinating systems~\cite{multicomponent1}.
Basic research in this area has enormously grown in the last few years~\cite{multicomponent2}. Due to the ability of optically trapping and cooling
gases to extremely low temperatures, it is possible to study different
phenomena in bosonic~\cite{Andrews,bosonicmixtures} as well as fermionic
mixtures~\cite{fermions}. Important quantum effects like Bose-Einstein
condensation (BEC) and superconductivity can now be studied in a very
controlled way in multicomponent atomic systems.

Interesting experiments with mixed bosonic quantum fluids have been done by
simultaneously trapping $^{87}Rb$ atoms in two different hyperfine states~\cite{Myatt,Hall-1, Hall-2, Lin}. The relative population is reached by
applying a coupling field characterized by a Rabi frequency $\Omega_R$ and a
detuning $\nu$ with respect to the spacing between the energy levels of the
two hyperfine states. In this way, it is possible to transfer atoms from one
hyperfine state to the other, producing a Josephson-type interaction between
species~\cite{Joseph-Exp.1,Legget-Sols-1,Legget-Sols-2}.

In general, the name \textquotedblleft Josephson
interaction\textquotedblright\ refers to the interaction of a large number
of bosonic degrees of freedom allowed to occupy two different quantum
states. Although it was originally proposed in superconductor systems~\cite
{Josephson}, where the bosons are Cooper pairs, there are many other systems
where this effect shows up. A review covering different physical systems can
be found in Ref.~[\cite{Barone}]. We can distinguish two types of
Josephson effects~\cite{Leggett-Review}: the so-called \textquotedblleft
external\textquotedblright , where the two states are spatially separated,
like, for instance, in BEC trapped in a double-well potential~\cite
{Albiez,doublewell,Gati,Levy}, or the \textquotedblleft
internal\textquotedblright , where the two bosonic states are
interpenetrated, without geometrical distinction, like, for instance, the
experiments in Refs.~\cite{Hall-1, Hall-2}. In this paper, we are mainly
interested in the latter case of internal Josephson-type interactions.

Static and dynamical properties of binary bosonic mixtures in different trap
geometries have been studied theoretically by essentially using
Gross-Pitaevskii equations~\cite{GPMixtures,Smerzi-1,Smerzi-2,Salerno,Bruno-1,Bruno-2,Bruno-3}.
Moreover, to study properties of uniform condensates, especially those
issues related with fluctuations, such as symmetry restoration, reentrances,
etc., quantum field theory at finite density and 
temperature~\cite{Rudnei-1,Rudnei-2,Rudnei-3,BarciFragaRamos}
is a useful technique. Related models, such as $O(N)$ models, have also been
extensively studied by using large-$N$ approximation and
renormalization-group techniques~\cite{largeN-1,largeN-2}. These papers are
mostly concentrated in multicomponent systems which conserve the particle
number of each species independently.

Motivated by these results, we decided to address the effect of
Josephson-like interactions in uniform bosonic mixtures. For simplicity, we
have considered an $O(2)$ model, perturbed with an explicit
symmetry-breaking term parametrized by the Rabi frequency $\Omega _{R}$ and
the detuning term $\nu $. This model is analyzed in mean-field approximation
plus Gaussian fluctuations. 

In the absence of Josephson interactions, this model is at the onset of  phase separation, since the two species are not physically distinguishable. However,  the presence  of Josephson interactions changes this scenario since it  explicitly breaks $O(2)$ symmetry. There is a  temperature regime where
the two atomic species uniformly condensate at the same critical temperature $T_{c}$
and their relative phase is locked by the phase of  the applied electromagnetic field responsible for the Rabi coupling and the  detuning. The relative population of each condensate strongly depends on the ratio $\Omega _{R}/|\nu |$. The main results of this paper are shown in Figures~(\ref{fig.nu})  and~(\ref{fig.Omega}) where we depict the condensate fraction of the two species as a function of temperature for different values of the parameter $\Omega _{R}/|\nu |$. Thus, controlling the external laser parameters,  {\em i.e.}, the Rabi coupling, the laser frequency (essentially the detuning) and 
the phase, it is possible to control each one of the condensate fractions as well as its phase difference. 
 
 An important result is that, due
to the original $O(2)$ symmetry, the effective Rabi frequency, given by 
$\Omega _{\mathrm{eff}}=\sqrt{\Omega _{R}^{2}+|\nu |^{2}}$ is strongly
renormalized by thermal fluctuations. On the other hand, the ratio $\Omega _{R}/|\nu |$, that controls the bosonic mixture, 
remains unaffected by quantum as well as thermal fluctuations. Thus,  the ratio between both condensates are temperature independent, allowing the possibility of control the relative condensate fractions with high accuracy.

The paper is organized as follows. In section \ref{QFT}, we describe a
general model for a binary mixture using quantum field theory language. In
section~\ref{O2}, we concentrate on the $O(2)$ model perturbed with
Josephson interactions. In section \ref{MF}, we present the mean-field
solution, while in section~\ref{Fluctuations} we analyze the effect of
fluctuations. Numerical results are presented in section \ref{Numerics} and,
finally, we discuss our results in section~\ref{discussion}. We reserve a brief appendix 
\ref{app} to describe the definitions of Rabi frequency and detuning parameter used to built our 
model. 

\section{A quantum field theory for binary bosonic mixtures}
\label{QFT} 

We will consider two bosonic species described by two complex fields, $\phi (\vec{x},t)$ and $\psi (\vec{x},t)$. The model is defined by the action 
\begin{equation}
S=\int d^{3}xdt\;\left\{ \mathcal{L}_{\psi }+\mathcal{L}_{\phi }+\mathcal{L}
_{I}\right\} ,  \label{S}
\end{equation}
where $\mathcal{L_{\psi }}$ and $\mathcal{L_{\phi }}$ are the
non-relativistic quadratic Lagrangian densities 
\begin{eqnarray}
\mathcal{L_{\psi }} &=&\psi ^{\ast }\left( i\partial _{t}+\frac{\nabla ^{2}}{
2m}+\mu _{\psi }\right) \psi \;,  \label{Lpsi} \\
\mathcal{L_{\phi }} &=&\phi ^{\ast }\left( i\partial _{t}+\frac{\nabla ^{2}}{
2m}+\mu _{\phi }\right) \phi \;.  \label{Lphi}
\end{eqnarray}
$\mu_\psi$ and $\mu_\phi$ are the chemical potentials for the $\psi$ and $\phi$ species, respectively.  We choose the same mass $m$ for both species, since
we are interested in mixtures composed by a single element in two different
hyperfine states.

It is convenient to parametrize the chemical potentials as 
\begin{eqnarray} 
\mu _{\phi }&=&\mu +\Omega _{R}
\label{muphi} \\
\mu _{\psi }&=&\mu -\Omega _{R}
\label{mupsi}
\end{eqnarray}
The parameter $\mu$ controls the overall particle
density at the time that the Rabi frequency $\Omega _{R}$ controls the
population imbalance (see Appendix~\ref{app} for the microscopic physical meaning 
of $\Omega_R$).
Throughout the paper, we have used a unit system in
which $\hbar =1$. 

The interaction Lagrangian density $\mathcal{L}_{I}$ can be split into two
terms, 
\begin{equation}
\mathcal{L}_{I}=\mathcal{L}_{c}+\mathcal{L}_{J}\;.  \label{LI}
\end{equation}
The first term, $\mathcal{L}_{c}$, contains two-body interactions that
preserve the particle number of each species individually. For diluted
gases, it can be approximated as a local quartic polynomial of the form 
\begin{equation}
\mathcal{L}_{c}=-\frac{g_{\psi }}{2}\left( \psi ^{\ast }\psi \right) ^{2}-
\frac{g_{\phi }}{2}\left( \phi ^{\ast }\phi \right) ^{2}-g_{\phi \psi }\psi
^{\ast }\psi \phi ^{\ast }\phi ,  \label{Lc}
\end{equation}
where the coupling constants $g_{\psi }=4\pi a_{\psi }/m$, $g_{\phi }=4\pi
a_{\phi }/m$ and $g_{\phi \psi }=8\pi a_{\phi \psi }/m$ are written in terms
of the intraspecies s-wave scattering lengths $a_{\psi }$, $a_{\phi }$ and
the interspecies s-wave scattering length $a_{\phi \psi }$. Note that this
interaction term is invariant under $U(1)_{\phi }\otimes U(1)_{\psi }$
transformations.

The second term of Eq.~(\ref{LI}) does not conserve the particle number of
each species individually. It conserves, however, the total particle number.
This term explicitly breaks the symmetry of Eq.~(\ref{Lc}) as $U(1)_{\phi
}\otimes U(1)_{\psi }\rightarrow U(1)_{\phi +\psi }$. We generally call
these terms as Josephson interactions, since they couple the phases of each
bosonic component. The simplest terms can be written as 
\begin{equation}
\mathcal{L}_{J}=\nu \psi ^{\ast }\phi +\nu ^{\ast }\phi ^{\ast }\psi -\frac{
g_{J}}{2}\left( \psi ^{\ast }\psi ^{\ast }\phi \phi +\phi ^{\ast }\phi
^{\ast }\psi \psi \right) .  \label{LJ}
\end{equation}
The quadratic term, proportional to $\nu $, and the quartic two-body
interaction term have, in general, very different origins. The one-particle
term is proportional to the detuning $\nu $, where 
we have considered a complex parameter in such a way to control the relative phases
of the condensates (see Appendix~\ref{app} for its definition). Considering the two species as components of an isospin
doublet, this term arises like an effective spin-orbit interaction~\cite{Garcia-March,DanWei}. 
We could also consider one-body terms of this type with derivative
couplings. However, to keep matters as simple as
possible, we will consider only this term. The second term in Eq.~(\ref{LJ})
represents scattering processes in which the internal hyperfine state of the
atoms is not conserved. In the absence of $\nu $, these processes are
unlikely to occur, since both hyperfine states are energetically well
separated. However, in the presence of a laser with small detuning between
the frequency differences, a very small coupling constant $g_{J}$ could
produce qualitatively different results.

Some aspects of the phase diagram of the model of Eqs.~(\ref{Lpsi}), (\ref{Lphi}) and~(\ref{Lc}), \emph{without Josephson couplings} ($\mathcal{L}_{J}=0$),
have been previously studied. The zero-temperature
mean-field analysis clearly establishes three different regimes, depending on relations between intra and inter species coupling constants.  If
\begin{equation}
g_{\phi }g_{\psi }-g_{\psi \phi }^{2}>0 \ ,
\label{constraint}
\end{equation}
it is possible to have  two coexisting condensates~\cite{Rudnei-1}.
Conversely, if 
\begin{equation}
g_{\phi }g_{\psi }-g_{\psi \phi }^{2}<0  
\label{constraint}
\end{equation}
both condensates cannot coexist and they
tend to spatially separate, producing an inhomogeneous state~\cite{exp-PS}. 
In addition, there is a special intermediate regime, 
\begin{equation}
g_{\phi }g_{\psi }-g_{\psi \phi }^{2}=0\; ,
\label{eq:intermediate}
\end{equation}  
that could be considered as the onset of homogeneous instability, since it is a fine tune region at the transition between the  homogeneous and the inhomogeneous ground states. 
Although it could be very difficult to experimentally reach this regime, it is a very interesting one due to its symmetry properties, as we will describe in the next section. 

\section{$O(2)$ model with Josephson anisotropy}
\label{O2} 

The model described in the preceding section has a very rich phase diagram
depending on the relative values of the coupling constants and of the
temperature. However, there is a special point of maximum symmetry where the
analysis gets simpler. Let us analyze model~(\ref{S}-\ref{LJ}) in its
maximum symmetry point given by $g_{\phi ,\psi }=g_{\phi }=g_{\psi }=g$, $\Omega_{R}=0$, $\nu =0$ and $g_{J}=0$. 
This point is at the intermediate regime described by Eq.~(\ref{eq:intermediate}).
The interaction term,
Eq.~(\ref{Lc}), takes the simpler form 
\begin{equation}
\mathcal{L}_{c}=-\frac{g}{2}\left( \psi ^{\ast }\psi +\phi ^{\ast }\phi
\right) ^{2}.  
\label{Lg}
\end{equation}
In addition to the $U(1)_{\phi }\otimes U(1)_{\psi }$ phase symmetry,
there is an emergent $O(2)$ symmetry, corresponding with rotations in the
isospin space $(\phi ,\psi )$. Thus, on the one hand, the particle number of each
species is independently conserved. On the other hand, the two species are physically
indistinguishable since any isospin rotation mixing the two species has
exactly the same action. Thus, the question of the difference between  homogeneous and  inhomogeneous  phases has no real meaning  at this point. However, an infinitesimal deviation of the coupling constants  leads the systems to one or to the other phase, depending on whether   $g_{\phi\psi}<g$ or $g_{\phi\psi}>g$. It is in this sense that we say that the $O(2)$  model,  $g_{\phi\psi}=g$,  is at the onset of the homogeneous instability. It is interesting to note that we  can rewrite the model  in terms of the real and imaginary components of the fields ($\Re\psi,\Im\psi, \Re\phi,\Im\phi$). In this representation, it is completely equivalent to a four-vector model  with $O(4)$ symmetry, which has been extensively studied in the literature related with the Chiral QCD phase transition~\cite{Wilczek} and, more recently, in the context of  Bose-Einstein condensates~\cite{Kleinert}.
 
Next, we minimally break the $O(2)$ symmetry unbalancing the chemical potentials 
with  a term proportional to $\Omega _{R}$ and a Josephson term of the form 
\begin{equation}
\mathcal{L}_{\nu }=\nu \psi ^{\ast }\phi +\nu ^{\ast }\phi ^{\ast }\psi .
\label{Lnu}
\end{equation}
For simplicity, we ignore two-body Josephson interactions (given by the term
proportional to $g_{J}$ in Eq.~(\ref{LJ})), since, in principle, it is of
higher order than the one-body interaction term we are considering.

The structure of this model is clearly visualized by defining new fields $(\varphi _{1},\varphi _{2})$ obtained by an isospin rotation of the original
fields $(\phi ,\psi )$, 
\begin{equation}
\left( 
\begin{array}{c}
\varphi _{1} \\ 
\varphi _{2}
\end{array}
\right) =M\left( 
\begin{array}{c}
\phi  \\ 
\psi 
\end{array}
\right) ,
\end{equation}
where the rotation matrix is 
\begin{equation}
M=\frac{1}{D}\left( 
\begin{array}{cc}
\Omega _{\mathrm{eff}}-\Omega _{R} & -\nu  \\ 
&  \\ 
\nu ^{\ast } & \Omega _{\mathrm{eff}}-\Omega _{R}
\end{array}
\right) ,  \label{M}
\end{equation}
with 
\begin{equation}
D=\sqrt{\left( \Omega _{\mathrm{eff}}-\Omega _{R}\right) ^{2}+|\nu |^{2}}\; , 
\end{equation}
and $\Omega _{\mathrm{eff}}=\sqrt{\Omega _{R}^{2}+|\nu |^{2}}$ is called the \emph{effective Rabi frequency} (see appendix~{\ref{app}). Of course,
one can immediately check that $\det (M)=1$. With this transformation, the
Lagrangian density takes the form 
\begin{eqnarray}
\mathcal{L} &=&\varphi _{1}^{\ast }\left( i\partial _{t}+\frac{\nabla ^{2}}{2m}+\mu _{+}\right) \varphi _{1}  +  \varphi _{2}^{\ast }\left( i\partial _{t}+\frac{\nabla ^{2}}{2m}+\mu_{-}\right) \varphi _{2} \nonumber \\
& & - \ \frac{g}{2}\left( \varphi _{1}^{\ast }\varphi _{1}+\varphi _{2}^{\ast}\varphi _{2}\right) ^{2},  
\label{Lvarphi}
\end{eqnarray}
where
\begin{eqnarray}
\mu _{+}&=&\mu +\Omega _{\mathrm{eff}},
\label{eq:mu+}\\
\mu _{-}&=&\mu -\Omega _{\mathrm{eff}}.
\label{eq:mu-}
\end{eqnarray} 
We see that, while
terms proportional to $\nu $ and $\Omega _{R}$  break the 
$O(2)$ and $U(1)_{\phi }\otimes U(1)_{\psi }$ symmetries, the system still
has an $U(1)_{\varphi _{1}}\otimes U(1)_{\varphi _{2}}$ symmetry in the new
variables. Thus, there is a direction in isospin space in which the particle
number of both species is still conserved independently. Equations~(\ref{eq:mu+}) and~(\ref{eq:mu-}), that define the chemical potentials in the new basis, are quite similar with 
Eqs.~(\ref{muphi}) and~(\ref{mupsi}) for the chemical potentials of $\psi$ and $\phi$,  with the difference that the Rabi frequency, $\Omega_R$ in the former case, should be substituted by the effective Rabi frequency, $\Omega_{\rm eff}$, in the latter. This simple behavior is a consequence of the $O(2)$ symmetry of the two-body interaction term, Eq.~(\ref{Lg}). It is not difficult to realize that, if we fix the coupling constants slightly away from the maximal symmetry point, $g_{\psi \phi }\neq g$, a term proportional to $\varphi _{1}\varphi _{1}\varphi _{2}^{\ast }\varphi _{2}^{\ast }$ would be
generated upon an isospin rotation,  breaking in this way $U(1)_{\varphi_{1}}\otimes U(1)_{\varphi _{2}}\rightarrow U(1)$. In this sense, the model
of Eq.~(\ref{Lvarphi}) implements a minimal perturbation of the complex $O(2)$ model.

Interestingly, Eq.~(\ref{Lvarphi}) does not depend on $\Omega _{R}$ and $\nu$ independently, but only depends on the effective Rabi frequency $\Omega _{\mathrm{eff}}=\sqrt{\Omega _{R}^{2}+|\nu |^{2}}$ (see Appendix~\ref{app} to see the relevance of the effective Rabi frequency in a simpler case of a two-level system). On the other hand, the
rotation matrix of Eq.~(\ref{M}) depends only on the ratio $\Omega _{R}/|\nu|$. It is instructive to see the form of the rotation matrix in two
different limits.

Let us consider, for instance, $|\nu |\ll \Omega _{R}$. In this case, 
\begin{equation}
M=\left( 
\begin{array}{cc}
\frac{|\nu |}{2\Omega _{R}} & -e^{i\alpha} \\ 
&  \\ 
e^{-i\alpha} & \frac{|\nu |}{2\Omega _{R}}
\end{array}
\right) ,
\end{equation}
where we have defined $\nu =|\nu |\exp (i\alpha)$. In the extreme
limit of $\nu \rightarrow 0$, both species are decoupled, as expected,
and the mixture is proportional to $\frac{|\nu |}{2\Omega _{R}}+O((|\nu|/2\Omega _{R})^{2})$.
In the opposite limit, $|\nu |\gg \Omega _{R}$, 
\begin{equation}
M=\frac{1}{\sqrt{2}}\left( 
\begin{array}{cc}
1-\frac{\Omega _{R}}{2|\nu |} & -e^{i\alpha}\left( 1+\frac{\Omega
_{R}}{2|\nu |}\right)  \\ 
&  \\ 
e^{-i\alpha}\left( 1+\frac{\Omega _{R}}{2|\nu |}\right)  & 1-\frac{\Omega _{R}}{2|\nu |}
\end{array}
\right) .
\end{equation}
In the extreme limit, $\Omega _{R}\rightarrow 0$, the fields are symmetrically
superposed, depending just on the phase of the detuning parameter, 
\begin{eqnarray}
\varphi _{1} &=&\frac{1}{\sqrt{2}}\left( \phi -e^{i\alpha}\psi\right) , \\
\varphi _{2} &=&\frac{1}{\sqrt{2}}\left( e^{-i\alpha}\phi +\psi\right) .
\end{eqnarray}
Small values of $\Omega _{R}$ produce corrections of order $\Omega _{R}/|\nu|$.

\section{Mean-field approximation}
\label{MF}

Let us analyze the model of Eq.~(\ref{Lvarphi}) in the mean-field
approximation. Minimizing the action $S=\int dtd^3x {\cal L}$ with ${\cal L}$ given by Eq.~(\ref{Lvarphi}), we obtain the equations of motion analogous to the  Gross-Pitaevskii equations 
\begin{eqnarray}
\left( i\partial _{t}+\frac{\nabla ^{2}}{2m}+\mu _{+}-g\left( \varphi
_{1}^{\ast }\varphi _{1}+\varphi _{2}^{\ast }\varphi _{2}\right) \right)
\varphi _{1} &=&0,
\label{meanfield1} \\
\left( i\partial _{t}+\frac{\nabla ^{2}}{2m}+\mu _{-}-g\left( \varphi
_{1}^{\ast }\varphi _{1}+\varphi _{2}^{\ast }\varphi _{2}\right) \right)
\varphi _{2} &=&0.
\label{meanfield2}
\end{eqnarray}
Looking for uniform and static solutions $\varphi _{1,2}(x,t)\equiv \varphi
_{1,2}^{0}$ we have 
\begin{eqnarray}
\left( \mu _{+}-g\left[ |\varphi _{1}^{0}|^{2}+|\varphi _{2}^{0}|^{2}\right]
\right) \varphi _{1}^{0} &=&0,  \label{MF1} \\
\left( \mu _{-}-g\left[ |\varphi _{1}^{0}|^{2}+|\varphi _{2}^{0}|^{2}\right]
\right) \varphi_{2}^{0} &=&0.  \label{MF2}
\end{eqnarray}
Assuming  that $\varphi_{1,2}^{0}\neq 0$, we can subtract Eq.~(\ref{MF2})
from Eq.~(\ref{MF1}), obtaining $\Delta \mu =\mu_{+}-\mu_{-}=0$.
Therefore, the two fields $\varphi _{1,2}$ cannot condensate simultaneously,
since a solution $\varphi_{1,2}^{0}\neq 0$ does not exist, except in the
case $\Delta \mu =2\Omega_{\mathrm{eff}}=0$. Instead, we have two possible
solutions, 
\begin{equation}
\varphi_{1}^{0}=0\;\;\;\;\;\;\;\;,\;\;\;\;\;\;\;|\varphi_{2}^{0}|^{2}=\mu_{-}/g
\end{equation}
or 
\begin{equation}
|\varphi_{1}^{0}|^{2}=\mu_{+}/g\;\;\;\;\;\;\;\;,\;\;\;\;\;\;\;\varphi_{2}^{0}=0 \ . 
\label{phi2=0} 
\end{equation}

Let us consider the solution $\varphi _{2}^{0}=0$, Eq.~(\ref{phi2=0}). Using
the matrix $M^{-1}$, given by the inverse of Eq.~(\ref{M}), it is simple to
turn back to the original fields, obtaining 
\begin{eqnarray}
\phi _{0} &=&\frac{\Omega_{\mathrm{eff}}-\Omega_{R}}{\sqrt{\left( \Omega_{\mathrm{eff}}-\Omega _{R}\right) ^{2}+|\nu |^{2}}}\;\varphi _{1}^{0}\;,
\label{phi0} \\
\psi _{0} &=&-\frac{\nu ^{\ast }}{\sqrt{\left( \Omega _{\mathrm{eff}}-\Omega_{R}\right) ^{2}+|\nu |^{2}}}\;\varphi _{1}^{0}\;,  
\label{psi0}
\end{eqnarray}
where $\phi_{0}$ and $\psi_{0}$ are the condensate amplitudes of the
fields $\phi (x)$ and $\psi (x)$, respectively. The first observation is
that the two original species $\phi $ and $\psi $ condense simultaneously
and the relative phase between these condensates, $\Delta \alpha $, is fixed
by the phase of the parameter $\nu $, 
\begin{equation}
\Delta \alpha =\alpha+\pi \ .
\end{equation}

At this point, it is important to emphasize this mean-field result. In the absence of 
Josephson interactions, the two species $\psi$ and $\phi$ 
cannot be distinguished from each other. In fact, the order parameter in this case is 
$|\varphi_1^0|^2+|\varphi_2^0|^2=|\psi_0|^2+|\phi_0|^2$, which is invariant under $O(2)$ transformations. 
The presence of Josephson interactions changes this situation since it breaks the $O(2)$ symmetry.
\begin{figure}[tbp]
\includegraphics[height=5.5 cm]{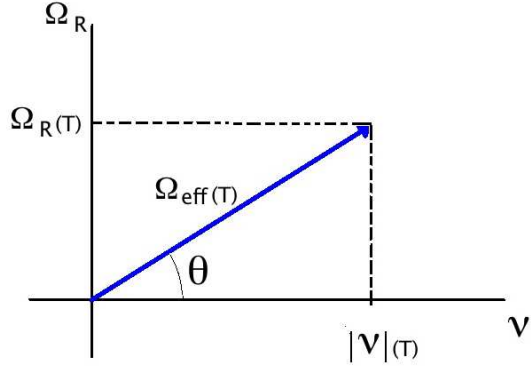}
\caption{Effective Rabi frequency $\Omega_{\rm eff}=\sqrt{\Omega_R^2+|\nu|^2}$ as parametrized by equations~(\ref{Omegatheta}) and~(\ref{nutheta}). While $\Omega_{\rm eff}$ is strongly temperature dependent, as shown in Eq.~(\ref{mu+mu-}), the angle given by $\tan\theta=\Omega_R/|\nu|$ is not affected by thermal fluctuations.}
\label{fig.Rabi}
\end{figure}
Moreover, the condensate fraction of both species depends on the ratio 
$\Omega _{R}/|\nu |$. It is instructive to parametrize $\Omega _{R}$ and 
$|\nu |$ in the following way (as shown in Fig.~(\ref{fig.Rabi})), 
\begin{eqnarray}
\Omega _{R} &=&\Omega_{\rm eff}\sin \theta , \label{Omegatheta} \\
|\nu | &=&\Omega_{\rm eff}\cos \theta ,
\label{nutheta}
\end{eqnarray}
with $0\leq \theta \leq \pi /2$. In terms of this parametrization, the ratio
between the condensate densities takes the form 
\begin{equation}
\frac{|\phi _{0}|^{2}}{|\psi _{0}|^{2}}=\sec ^{2}\theta \left( 1-\sin \theta
\right) ^{2},  \label{phi/psi}
\end{equation}
which does not depend on $\Omega_{\rm eff} $ but only on $\tan \theta =\Omega
_{R}/|\nu |$. We depict this function in Fig.~(\ref{fig.theta}). 
\begin{figure}[ht]
\includegraphics[height=5.5 cm]{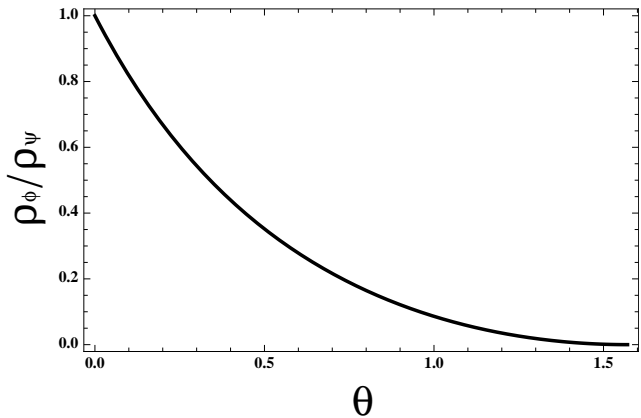}
\caption{$|\protect\phi _{0}|^{2}/|\protect\psi _{0}|^{2}$ given by 
Eq.~(\protect\ref{phi/psi}) as a function of $\protect\theta $, where $\tan\theta=\Omega_R/|\nu|$.}
\label{fig.theta}
\end{figure}
For $\theta \rightarrow 0$ or $\Omega _{R}\rightarrow 0$ with $|\nu |\neq 0$, both condensates have essentially the same fraction. On the other hand, for $\theta \rightarrow \pi /2$ or $|\nu |\rightarrow 0$ with $\Omega
_{R}\neq 0$, only one of the fields condensates. We will show in the next section that, while $\Omega_{\rm eff}$ is renormalized by temperature, the present result is temperature independent.

\section{Effect of Fluctuations}
\label{Fluctuations}

To study thermal as well as quantum fluctuations, we start by considering
the following Euclidean ($\tau=i t$) finite temperature field theory: 
\begin{eqnarray}
& & S_{\mathrm{E}}(\beta ) =\int_{0}^{\beta }d\tau \int d^{3}x\left[ \varphi_{1}^{\ast }\left( \partial _{\tau }-\frac{\nabla ^{2}}{2m}-\mu _{+}\right) \varphi _{1}\right.   \nonumber \\
&+& \left.  \varphi _{2}^{\ast }\left( \partial _{\tau }-\frac{\nabla ^{2}}{2m}-\mu_{-}\right) \varphi _{2}  + \frac{g}{2}\left( \varphi _{1}^{\ast }\varphi _{1}+\varphi_{2}^{\ast }\varphi _{2}\right) ^{2}\right] 
\label{SE}
\end{eqnarray}
with $\beta =1/T$. The partition function reads 
\begin{eqnarray}
Z(\beta ,\vec{J}) &=&\int \mathcal{D}\varphi _{1}\mathcal{D}\varphi_{1}^{\ast }\mathcal{D}\varphi _{2}\mathcal{D}\varphi _{2}^{\ast }\;e^{-S_{\mathrm{E}}+\int d^{3}xd\tau \vec{J}\cdot \vec{\varphi}}  \nonumber \\
&=&e^{-\beta VW[\beta ,J]},  \label{Z}
\end{eqnarray}
where we have introduced a source $\vec{J}$ in order to compute field
correlation functions. The functional integration measure implicitly
contains the cyclic bosonic boundary condition in Euclidean time, $\varphi
_{1,2}(0,x)=\varphi _{1,2}(\beta ,x)$. $W[\beta ,\vec{J}]=-\frac{1}{\beta V}
\ln Z$ is the Helmholtz free energy density.

The main purpose of this section is to compute $W[\beta ,J]$ in mean-field
approximation plus Gaussian fluctuations. 
We expect that, at least in a certain temperature range, fluctuations will not change the general mean-field structure. With
this in mind, in order to compute $W[\beta ,J]$ we replace in Eq.~(\ref{Z})
the following decomposition 
\begin{eqnarray}
\varphi _{1}(x,\tau ) &=&\varphi _{1}^{0}+\tilde{\varphi}_{1}(x,\tau ) \label{decomposition1} \\
\varphi _{2}(x,\tau ) &=&\tilde{\varphi}_{2}(x,\tau ) \label{decomposition2}
\end{eqnarray}
in which $\int d^{3}x\;\tilde{\varphi}_{1,2}=0$ and $\varphi _{1}^{0}(J)$ is
a solution of the mean field equations, 
\begin{eqnarray}
\left. \frac{\delta S_{E}}{\delta \varphi _{1}}\right\vert _{\varphi_{1}=\varphi _{1}^{0},\varphi _{2}=0} &=&J\;, \\
\left. \frac{\delta S_{E}}{\delta \varphi _{2}}\right\vert _{\varphi_{1}=\varphi _{1}^{0},\varphi _{2}=0} &=&0\;,
\end{eqnarray}
where we have chosen a constant source $\vec{J}$, pointing in the $\varphi _{1}$ direction.

Retaining up to second-order terms in the fluctuations we obtain 
\begin{equation}
Z(\beta )=e^{-\beta VU_{0}(\varphi _{1}^{0})}\int \left[ \mathcal{D}\tilde{\varphi}\right] \;e^{-\int d\tau d^{3}x\sum_{ij}\tilde{\varphi}_{i}^{\ast}S_{ij}^{(2)}\tilde{\varphi}_{j}}\;,
\end{equation}
where 
\begin{equation}
U_{0}=-\mu _{+}|\varphi _{1}^{0}|^{2}+\frac{g}{2}|\varphi _{1}^{0}|^{4}\;\;.
\end{equation}
The integration measure is 
\begin{equation}
\lbrack \mathcal{D}\tilde{\varphi}]=\mathcal{D}\tilde{\varphi}_{1}\mathcal{D} \tilde{\varphi}_{1}^{\ast }\mathcal{D}\tilde{\varphi}_{2}\mathcal{D}\tilde{\varphi}_{2}^{\ast}
\end{equation}
and the quadratic kernel,
\begin{equation}
S_{ij}^{(2)}=\left. \frac{\delta ^{2}S_{\mathrm{E}}}{\delta \varphi_{j}^{\ast}\delta\varphi _{i}}\right\vert _{\varphi _{1}=\varphi_{1}^{0},\varphi _{2}=0}\;\;,
\end{equation}
with $i,j=1,2$.

Integrating out quadratic fluctuations, we find an expression for the free
energy density, 
\begin{equation}
W[J,\beta ]=U_{0}+\Delta W  \label{W}
\end{equation}
with 
\begin{equation}
\Delta W[J,\beta ]=\frac{1}{2}\ln \det \hat{S}^{(2)}=\frac{1}{2}{\rm Tr}\ln \hat{S}^{(2)}\;.  \label{DeltaW}
\end{equation}

The matrix $\hat{S}^{(2)}$ in the \{Re$(\tilde{\varphi}_{1})$, Im$(\tilde{\varphi}_{1})$, Re$(\tilde{\varphi}_{2})$, Im$(\tilde{\varphi}_{2})$\} basis
decouples into two independent $2\times 2$ blocks, 
\begin{equation}
\hat{S}^{(2)}=\left( 
\begin{array}{cc}
\hat{S}_{a}^{(2)} & 0 \\ 
0 & \hat{S}_{b}^{(2)}
\end{array}
\right) ,
\end{equation}
with 
\begin{equation}
\hat{S}_{a}^{(2)}=\left( 
\begin{array}{cc}
-\frac{\nabla ^{2}}{2m}-\mu _{+}+3g|\varphi _{1}^{0}|^{2} & i\partial _{\tau} \\ 
-i\partial _{\tau } & -\frac{\nabla ^{2}}{2m}-\mu _{+}+g|\varphi_{1}^{0}|^{2}
\end{array}
\right) 
\end{equation}
and 
\begin{equation}
\hat{S}_{b}^{(2)}=\left( 
\begin{array}{cc}
-\frac{\nabla ^{2}}{2m}-\mu _{-}+g|\varphi _{1}^{0}|^{2} & i\partial_{\tau } \\ 
-i\partial _{\tau } & -\frac{\nabla ^{2}}{2m}-\mu _{-}+g|\varphi_{1}^{0}|^{2}
\end{array}
\right) .
\end{equation}

It is not difficult to compute the trace in Fourier space, obtaining 
\begin{equation}
\Delta W=\frac{1}{2\beta }\sum_{n=-\infty }^{+\infty }\int \frac{d^{3}q}{(2\pi )^{3}}\ln \left\{ \left( \omega _{n}^{2}+E_{1}^{2}\right) \left(\omega _{n}^{2}+E_{2}^{2}\right) \right\} ,
\end{equation}
where $\omega _{n}=2\pi n/\beta $ are the Matsubara frequencies, 
\begin{equation}
E_{1}=\sqrt{\left( \frac{q^{2}}{2m}-\mu _{+}+3g|\varphi _{1}^{0}|^{2}\right)
\left( \frac{q^{2}}{2m}-\mu _{+}+g|\varphi _{1}^{0}|^{2}\right) }  \label{E1}
\end{equation}
and 
\begin{equation}
E_{2}=\frac{q^{2}}{2m}-\mu _{-}+g|\varphi _{1}^{0}|^{2}\;.  \label{E2}
\end{equation}
Summing up the Matsubara frequencies, using 
\begin{equation}
\frac{1}{\beta }\sum_{n}\ln (\omega _{n}^{2}+E_{i}^{2})=E_{i}+\frac{2}{\beta}\ln \left( 1-e^{-\beta E_{i}}\right) ,
\end{equation}
we obtain 
\begin{equation}
\Delta W=\frac{1}{2}\int \frac{d^{3}q}{(2\pi )^{3}}\sum_{i}\left\{ E_{i}+
\frac{2}{\beta }\ln \left( 1-e^{-\beta E_{i}}\right) \right\} .
\end{equation}

It is interesting to note that, if we substitute the mean-field value for $\varphi _{1}^{0}$, given by Eq.~(\ref{phi2=0}), into Eqs.~(\ref{E1}) and~(\ref{E2}), we immediately obtain 
\begin{equation}
\tilde{E}_{1}=\sqrt{\left( \frac{q^{2}}{2m}\right) \left( \frac{q^{2}}{2m} + 2g|\varphi _{1}^{0}|^{2}\right) }  
\label{Goldstone}
\end{equation}
and 
\begin{equation}
\tilde{E}_{2}=\frac{q^{2}}{2m}+2 \Omega _{\rm eff}\;.  
\label{Gap}
\end{equation}
Equations.~(\ref{Goldstone}) and~(\ref{Gap}) are the usual energy excitations
computed in the Bogoliubov approximation. Note that $\lim_{q\rightarrow 0}
\tilde{E}_{1}=0$, corresponding with the Goldstone mode associated with the
spontaneous breakdown of the  $U_{\varphi _{1}}(1)$ symmetry, while Eq.~(\ref{Gap}) is a gapped mode corresponding to non-condensate fluctuations.

It is useful to express the free energy $W(\beta ,J)$ in terms of the order parameter: 
\begin{equation}
\bar{\varphi}=\delta W/\delta J=\varphi _{1}^{0}+\frac{1}{2}\mathrm{Tr}\left[\frac{1}{\hat{S}^{(2)}}\frac{\delta \hat{S}^{(2)}}{\delta \varphi _{1}^{0}} \frac{\delta \varphi _{1}^{0}}{\delta J}\right] \;.  
\label{OP}
\end{equation}
At mean-field level, the order parameter is exactly the mean-field solution $\varphi _{1}^{0}$. However, when fluctuations are taken into account, the result given by Eq.~(\ref{OP}) is more involved.

We define the Gibbs free energy as a functional of the order parameter $\bar\varphi$ by making a Legendre transformation 
\begin{equation}
\Gamma[\beta, \bar\varphi]= \bar\varphi J- W \; ,  \label{Gibbs}
\end{equation}
where $\delta \Gamma/\delta\bar\varphi=J$. In Eq.~(\ref{Gibbs}), $J$ is a
function of the order parameter $\bar\varphi$ obtained by inverting Eq.~(\ref{OP}). To leading order in the fluctuations the result is 
\begin{eqnarray}
\Gamma[\beta, \bar\varphi]&=& \mu_+ |\bar\varphi|^2-\frac{g}{2}|\bar\varphi|^4  \nonumber \\
&-&\frac{1}{2}\int \frac{d^3q}{(2\pi)^3} \sum_i \left\{E_i+\frac{2}{\beta}\ln\left(1-e^{-\beta E_i} \right) \right\}.  
\label{Gamma}
\end{eqnarray}
This is the Gibbs free energy computed at mean field plus Gaussian
fluctuations or, in the language of quantum field theory, the finite
temperature one-loop effective action.

The actual condensate amplitude $\bar{\varphi}_{m}$ is computed by
minimizing the free energy, 
\begin{equation}
\left. \frac{\partial \Gamma \lbrack \beta ,\bar{\varphi}]}{\partial \bar{\varphi}}\right\vert _{\bar{\varphi}=\bar{\varphi}_{m}}=0.
\label{minimizing}
\end{equation}
By analogy with the mean field solution $|\varphi_1^0|^2=\mu_+/g$ we can define an 
effective chemical potential in the following way, 
\begin{equation}
|\bar{\varphi}_{m}|^{2}(T)=\frac{1}{g}\;\bar\mu_{+}(T),
\label{eq:barmu+}
\end{equation}
where now $\bar\mu_+(T)$ is the effective chemical potential for the $\varphi_1$ component, renormalized by quantum as well as thermal fluctuations.
Using Eqs.~(\ref{Gamma}) and~(\ref{minimizing}), we obtain an expression for $\bar{\mu}_{+}(T)$
in terms of the original bare $\mu _{+}$, 
\begin{eqnarray}
\lefteqn{\bar{\mu}_{+}=\mu _{+}-\frac{1}{2}\int \frac{d^{3}q}{(2\pi )^{3}} \times }  \nonumber \\
&\times &\left\{ \frac{(2\frac{q^{2}}{2m}+\bar{\mu}_{+})(1+2n(E_{1}))}{\sqrt{\left( \frac{q^{2}}{2m}\right) \left( \frac{q^{2}}{2m}+2\bar{\mu}_{+}\right)}}+1+2n(E_{2})\right\} ,  
\label{mu+}
\end{eqnarray}
where $n(E_{i})$ is the usual Bose distribution 
\begin{equation}
n(E_{i})=\frac{1}{e^{\beta E_{i}}-1}  \label{Bose}
\end{equation}
with $i=1,2$.

The total particle density of each species can be computed as 
\begin{eqnarray}
\rho_{\varphi_1}&=&\left.\frac{\partial \Gamma}{\partial\mu_+}\right|_{\bar\varphi_m}= \frac{\mu_+}{g}-\frac{1}{2} \int \frac{d^3q}{(2\pi)^3} \frac{q^2/2m}{E_1} (1+2n(E_1))  \nonumber \\
&&\mbox{\hspace*{1.8 cm}}-\frac{1}{2}\int \frac{d^3q}{(2\pi)^3} (1+2n(E_2)) \;, \\
\rho_{\varphi_2}&=&\left.\frac{\partial \Gamma}{\partial\mu_-}
\right|_{\bar\varphi_m}=\frac{1}{2}\int \frac{d^3q}{(2\pi)^3} (1+2n(E_2))\;.
\end{eqnarray}
Using the relation between $\mu_+$ and $\bar\mu_+$ given by Eq.~(\ref{mu+}),
we finally get 
\begin{eqnarray}
\rho_{\varphi_1}&=& \frac{\bar\mu_+}{g}+\frac{1}{2} \int \frac{d^3q}{(2\pi)^3} \frac{\frac{q^2}{2m}+\bar\mu_+}{E_+}\coth(\beta E_+/2),  \label{rho1} \\
\rho_{\varphi_2}&=&\frac{1}{2}\int \frac{d^3q}{(2\pi)^3} \coth(\beta E_-/2),
\label{rho2}
\end{eqnarray}
with 
\begin{eqnarray}
E_+&=&\sqrt{\left(\frac{q^2}{2m}\right)\left(\frac{q^2}{2m}+2\bar\mu_+(T)
\right)},  \label{E+} \\
E_-&=&\frac{q^2}{2m}+2\bar\Omega_{\rm eff}(T)\;,  
\label{E-}
\end{eqnarray}
where we have defined the renormalized effective Rabi frequency $\bar\Omega_{\rm eff}(T)=(\bar\mu_+-\bar\mu_-)/2 $ as a difference between the renormalized chemical potentials, in analogy with  the bare effective Rabi frequency $\Omega_{\rm eff}=(\mu_+-\mu_-)/2 $.    Notice that, while Eq.~(\ref{rho1}) completely determines $\bar\mu_+$, Eq.~(\ref{rho2}) is the definition of the renormalized chemical potential $\bar\mu_-$, through the expression for the excitation energy $E_-$ (Eq.~(\ref{E-})). In terms of these variables, Eqs.~(\ref{rho1}) and~(\ref{rho2}) are coupled equations. However, it is more convenient to work with $\bar\mu_+$ and $\bar\Omega_{\rm eff}$ as independent variables, in such a way that  Eqs.~(\ref{rho1}) and~(\ref{rho2}) are now decoupled equations. In terms of these variables, all other  chemical potentials are linear combinations of the former, such as,  $\bar\mu_-=\bar\mu_+-2\bar\Omega_{\rm eff}$ and $\bar\mu=\bar\mu_+-\bar\Omega_{\rm eff}$.

Expressions~(\ref{rho1}) and~(\ref{rho2}) have the usual ultraviolet
divergences of a field theory at $T=0$. As is well known, temperature
fluctuations are always convergent. The usual way to deal with this
divergence is to regularize the integral and then renormalize the bare
constants $\mu _{+}$, $\mu _{-}$ and $g$, in order to obtain finite
results. A convenient procedure, in the non-relativistic scalar case, is the
cut-off technique. If we simply limit the momentum integrals using an
ultraviolet cut-off, $0\leq |\vec{q}|\leq \Lambda $, the results are
obviously $\Lambda $-dependent. However, if we begin the calculations with
renormalized constants, $\mu _{\pm }^{R}=\mu _{\pm }+\delta \mu _{\pm
}(\Lambda )$, we can adjust $\delta \mu _{\pm }(\Lambda )$ to make the
result independent of $\Lambda$. At the end, we can safely take the limit $\Lambda \rightarrow \infty $. After this procedure, the renormalized expressions read 
\begin{eqnarray}
\rho _{\varphi _{1}} &=&\frac{\bar{\mu}_{+}}{g}+\frac{(m\bar{\mu}_{+})^{3/2}}{3\pi ^{2}}+\int \frac{d^{3}q}{(2\pi )^{3}}\frac{\frac{q^{2}}{2m}+\bar{\mu}_{+}}{E_{+}(e^{\beta E_{+}}-1)},  \label{rho1R} \\
\rho _{\varphi _{2}} &=&\int \frac{d^{3}q}{(2\pi )^{3}}\frac{1}{e^{\beta E_{-}}-1} \ .  
\label{rho2R}
\end{eqnarray}
Equation~(\ref{rho1R}) implicitly defines the condensate density $\bar{\varphi}_{m}(T)$
 or, equivalently, the effective chemical potential $\bar{\mu}_{+}(T)$, given by Eq.~(\ref{eq:barmu+}). This equation coincides with that derived from a one-loop effective
potential of a single self-interacting field~\cite{Rudnei-1}. Moreover, 
eq.~(\ref{rho2R}) determines the effective Rabi frequency, 
$\bar\Omega _{\mathrm{eff}}(T)$ through the expression for $E_-$, Eq.~(\ref{E-}).     
In Eqs.~(\ref{rho1R}) and~(\ref{rho2R}), $\rho_{\varphi_1}$ and $\rho_{\varphi_2}$ are two independent constants, since the particle number of each species is conserved independently, due to the symmetry 
$U_{\varphi_1}(1)\otimes U_{\varphi_2}(1)$. 
The critical
temperature, $T_{c}$, is easily computed by fixing $\bar{\mu}_{+}(T_c)=0$ in 
Eq.~(\ref{rho1R}), obtaining the usual expression for an ideal gas, 
\begin{equation}
T_{c}=\frac{2\pi }{m\zeta (3/2)^{2/3}}\rho _{\varphi _{1}}^{2/3},  \label{Tc}
\end{equation}
with $\zeta (3/2)\sim 2.612$. We expect corrections of $T_{c}$ only at a
two-loop approximation~\cite{Tc1,Tc2}. 
Since $E_-$ are gapped energy excitations, the integral in  Eq.~(\ref{rho2R}) can be safely done in the classical limit. Solving  for $\bar\Omega_{\rm eff}(T)$ we obtain, 
\begin{equation}
\bar\Omega _{\mathrm{eff}}(T)=\frac{T}{2}\ln \left[ \left( \frac{\rho
_{\varphi _{1}}}{\rho _{\varphi _{2}}}\right) \left( \frac{T}{T_{c}}\right)
^{3/2}\right] \;.  \label{mu+mu-}
\end{equation}
Note that there is a minimum temperature for which $\bar\Omega_{\rm eff}(T_r)=0$, given by 
$T_{r}=(\rho _{\varphi_{2}}/\rho _{\varphi _{1}})^{2/3}T_{c}$.
At this temperature, the $O(2)$ symmetry is restored. This reentrance transition makes the excitation energy $E_-$ (Eq.~(\ref{E-})) gapless, producing an instability of  the mean-field solution. Then, at this temperature, the chosen mean-field solution is unstable under Gaussian fluctuations.  
In order to have the condensate structure given by Eqs.~(\ref{phi0}) and~(\ref{psi0}), we need to
fix $\rho _{\varphi _{2}}/\rho _{\varphi _{1}}<1$ and $T_{r}<T<T_{c}$. 
In the next section, we numerically compute the condensate fractions as
functions of temperature for different values of the parameters.

\section{Numerical results}
\label{Numerics}

To compute the condensate density profile we rewrite Eq.~(\ref{rho1R}) in
dimensionless form. For this, we define the condensate fraction $\rho _{c}=\bar{\mu}_{+}/(g\rho _{\varphi _{1}})$. The dimensionless temperature is
defined as $\bar{T}=T/T_{c}$ and we introduce the diluteness parameter $n_{\varphi _{1}}=\rho _{\varphi _{1}}a^{3}$, where $a$ is the s-wave
scattering length. Using these definitions, we can write Eq.~(\ref{rho1R})
in the following form: 
\begin{eqnarray}
1 &=&\rho _{c}+\frac{8}{3\pi ^{1/2}}n_{\varphi _{1}}^{1/2}\rho _{c}^{3/2} \nonumber \\
&+&\frac{4}{\pi ^{1/2}\zeta (3/2)}\bar{T}^{3/2}\int_{0}^{\infty }dyy\frac{y^{2}+2\zeta (3/2)^{2/3}n_{\varphi _{1}}^{1/3}\rho _{c}\bar{T}^{-1}}{\sqrt{y^{2}+4\zeta (3/2)^{2/3}n_{\varphi _{1}}^{1/3}\rho _{c}\bar{T}^{-1}}} \nonumber \\
&\times &\left( e^{y\sqrt{y^{2}+4\zeta (3/2)^{2/3}n_{\varphi _{1}}^{1/3}\rho_{c}\bar{T}^{-1}}}-1\right) ^{-1}\;\;.  
\label{rhoca}
\end{eqnarray}
It is simple to check that the limit $n_{\varphi _{1}}\rightarrow 0$ leads
to the ideal gas result $\rho _{c}=1-\bar{T}^{3/2}$. The second term of the
r.h.s. of eq.~(\ref{rhoca}) gives the quantum depletion of the condensate,
while the third term represents the temperature dependence. Numerically
solving Eq.~(\ref{rhoca}), we can obtain the condensate fraction $\rho _{c}(\bar{T})$ for different values of the diluteness parameter $n_{\varphi _{1}}$. 
From this result, it is simple to compute the condensate fractions for the
original fields $\phi $ and $\psi $, using Eqs.~(\ref{phi0}) and~(\ref{psi0}).

We define the condensate fractions for the fields $\phi$ and $\psi$ as 
$\rho^c_{\phi}=|\phi_0|^2/(\rho_{\varphi_1}+\rho_{\varphi_2})$ and 
$\rho^c_{\psi}=|\psi_0|^2/(\rho_{\varphi_1}+\rho_{\varphi_2})$, where we
chose the total particle density $\rho_{\phi}+\rho_{\psi}=\rho_{\varphi_1} + \rho_{\varphi_2}$ to normalize the fractions. Then, we use Eqs.~(\ref{phi0})
and~(\ref{psi0}) to relate $\rho^c_{\phi}$ and $\rho^c_{\psi}$ with $\rho_c$, given by Eq.~(\ref{rhoca}).

There are two interesting regimes to focus on. For $|\nu |/\Omega _{R}\ll 1$, the condensate fractions become 
\begin{eqnarray}
\rho _{\phi }^{c} &\sim &\left( 1-\frac{\rho _{\varphi _{2}}}{\rho _{\varphi
_{1}}}\right) \left( \frac{|\nu |}{2\Omega _{R}}\right) ^{2}\rho _{c}\;, \label{numinorOmega1} \\
\rho _{\psi }^{c} &\sim &\left( 1-\frac{\rho _{\varphi _{2}}}{\rho _{\varphi
_{1}}}\right) \rho _{c}\;.  
\label{numinorOmega2}
\end{eqnarray}
The first factor compensates the normalizations of $\rho _{\phi ,\psi }^{c}$
and $\rho _{c}$. To obtain it, we have considered $\rho _{\varphi _{2}}/\rho
_{\varphi _{1}}<1$ and we have dropped terms proportional to $(\rho
_{\varphi _{2}}/\rho _{\varphi _{1}})^{2}$. The condensate fraction is
determined by the factor $(|\nu |/2\Omega _{R})^{2}$ and the next
corrections to eqs.~(\ref{numinorOmega1}) and (\ref{numinorOmega2}) are
proportional to $(|\nu |/2\Omega _{R})^{4}$. In Fig.~(\ref{fig.nu}) we
show the typical profile of both condensates, where we have fixed $n_{\varphi _{1}}=10^{-5}$, 
$\rho _{\varphi _{2}}/\rho_{\varphi _{1}}=10^{-1}$ and $|\nu |/2\Omega _{R}=0.24$. Note that $\rho _{\phi }^{c}$ is strongly
suppressed by the factor $\nu /\Omega _{R}$ and tends to disappear in the
limit $|\nu |\rightarrow 0$. An interesting observation is that the factor $|\nu |/\Omega _{R}$ is not corrected by temperature fluctuations. This is a
direct consequence of the $O(2)$ symmetry of the two-body interaction. 
\begin{figure}[tbp]
\includegraphics[height=6 cm]{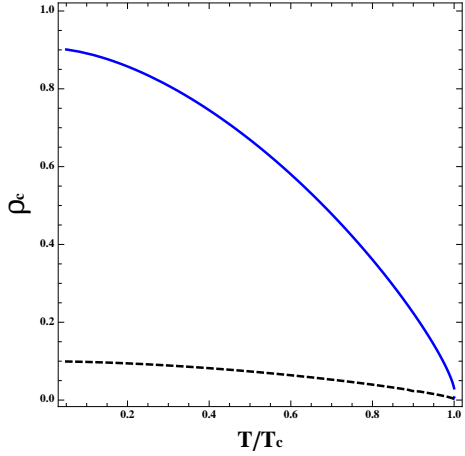}
\caption{Condensate fractions as functions of the dimensionless temperature $\bar{T}$ in the limit $|\protect\nu |/\Omega _{R}<1$. The solid line
represents $\protect\rho _{\protect\psi }^{c}$, given by eq.~(\protect\ref{numinorOmega2}), while the dashed line is $\protect\rho _{\protect\phi }^{c}$, given by eq. (\protect\ref{numinorOmega1}). We have fixed 
$n_{\protect\varphi _{1}}=10^{-5}$, $\protect\rho _{\protect\varphi _{2}}/\protect\rho _{\protect\varphi _{1}}=10^{-1}$ 
and $|\protect\nu |/2\Omega _{R}=0.24$.}
\label{fig.nu}
\end{figure}

In the opposite regime $\Omega_R/ |\nu|\ll1$, the condensate densities of
both species are essentially equal, with small corrections, given by 
\begin{eqnarray}  \label{Omegaminornu1}
\rho^c_{\phi_0}&\sim& \frac{1}{2}\left(1-\frac{\rho_{\varphi_2}}{\rho_{\varphi_1}}\right) \left(1-\frac{\Omega_R}{|\nu|}\right) \rho_c \; , \\
\rho^c_{\psi_0}&\sim& \frac{1}{2}\left(1-\frac{\rho_{\varphi_2}}{\rho_{\varphi_1}}\right) \left(1+\frac{\Omega_R}{|\nu|}\right) \rho_c \;,
\label{Omegaminornu2}
\end{eqnarray}
where we have discarded corrections of order $(\Omega_R/|\nu|)^2$. We show
these curves in Fig.~(\ref{fig.Omega}) for $n_{\varphi_1}=10^{-5}$, $\rho_{\varphi_2}/\rho_{\varphi_1}=10^{-1}$ and $\Omega_R/|\nu|=0.02$.

\begin{figure}[tbp]
\includegraphics[height=6 cm]{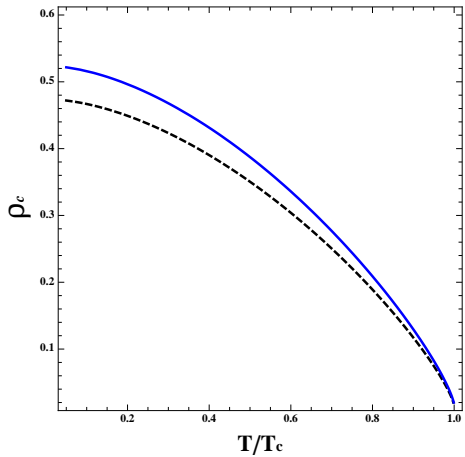}
\caption{Condensate fractions as functions of the dimensionless temperature $\bar{T}$ in the limit $\Omega _{R}/|\protect\nu |<1$. The solid line represents $\protect\rho _{\protect\psi }^{c}$, given by eq.~(\protect\ref{Omegaminornu2}), while the dashed line is $\protect\rho _{\protect\phi }^{c}$, given by eq. (\protect\ref{Omegaminornu1}). We have fixed $n_{\protect\varphi _{1}}=10^{-5}$, $\protect\rho _{\protect\varphi _{2}}/\protect\rho _{\protect\varphi _{1}}=10^{-1}$ and $\Omega _{R}/|\protect\nu |=0.02$.}
\label{fig.Omega}
\end{figure}

\section{Discussion}
\label{discussion} 

We have addressed the problem of equilibrium properties of a uniform mixture
of two bosonic fields in the presence of Josephson-type interactions. We
have considered a quantum field theory built by two non-relativistic complex
bosonic fields with general two-body local interactions. We have focused on
a particular symmetry point, in which, in addition to the $U(1)\otimes U(1)$
phase symmetry, there is an emergent $O(2)$ symmetry, related with
rotations in the isospin space $(\phi ,\psi )$. 
We have minimally perturbed this model by considering the effect of
Josephson couplings that unbalance the species population by transferring
charge from one species to the other. These interactions are parametrized by
the Rabi frequency $\Omega _{R}$ and the detuning $\nu $. By making a
rotation in the isospin space, $(\phi ,\psi )\rightarrow (\varphi
_{1},\varphi _{2})$, we have shown that there is a special direction for
which the $U(1)\otimes U(1)$ phase symmetry is recovered and only one of the
bosonic species (say $\varphi _{1}$) could eventually condensate in this
framework. In this basis, the density of each bosonic species $\rho
_{\varphi _{1}}$ and $\rho _{\varphi _{2}}$ is conserved independently. Of
course, the $O(2)$ symmetry is still broken, provided the difference between
chemical potentials $\Delta \mu =\mu _{+}-\mu _{-}=2\Omega _{\mathrm{eff}}\neq 0$.

In the $(\varphi _{1},\varphi _{2})$ basis, it is simpler to compute
fluctuations. Specifically, we have computed finite temperature one-loop
effective action (the Gibbs free energy) as a function of the order
parameter and the temperature. In this way, by minimizing the free
energy, we have obtained the condensate fraction. Since the total density of each 
species is conserved in this basis, the constant values of $\rho_{\varphi_1}$ and $\rho_{\varphi_2}$ completely determine the two chemical potentials $\bar\mu_+$ and $\bar\mu_-$. Alternatively, there is an interesting decoupling if we work in terms of the parameters $\bar\mu_+$ and $\bar\Omega_{\rm eff}$.  While the density $\rho_{\varphi_1}$ fixes the value of 
$\bar\mu_+(T)$, the value of $\rho_{\varphi_2}$ determines the value of $\bar\Omega_{\rm eff}(T)$. In this way, we can explicitly compute two limiting temperatures given by $\bar\mu_+(T_c)=0$ and 
$\Omega_{\rm eff}(T_r)=0$.  $T_c$ is the critical temperature for the Bose-Einstein condensation and $T_r$ is a reentrance temperature where the $O(2)$ symmetry is recovered.  Below this temperature, the
mean-field solution is unstable under thermal fluctuations. Thus, our
results are only valid for $T_{r}<T<T_{c}$. To compute the condensate fractions below $T_r$, it is necessary to assume that both species in the rotated frame ($\varphi_{1}, \varphi_{2}$) could condensate,  making the computation of fluctuations more involved.

To obtain the condensate profiles of the original fields, we rotated back to
the original basis $(\phi ,\psi )$. This rotation only depends on the ratio $\Omega _{R}/|\nu |$. It is interesting to note that, due to the $O(2)$
symmetry of the two-body interaction, fluctuations only renormalize the
effective Rabi frequency $\Omega _{\mathrm{eff}}=\sqrt{\Omega _{R}^{2}+|\nu
|^{2}}$, while the ratio $\Omega _{R}/|\nu |$ remains unaffected. Thus, the
isospin rotation coefficients are temperature independent.

In figures~(\ref{fig.nu}) and~(\ref{fig.Omega}) we show the condensate
profiles of the $\psi $ and $\phi $ species as functions of the temperature
for different values of the parameter $\Omega _{R}/|\nu |$. We have shown
that, for a temperature interval $T_{r}<T<T_{c}$, both bosonic species
condensate and the relatives phases are locked by the laser  phase $\alpha$. We also have shown that the ratio between the condensates
essentially depends on the temperature-independent parameter $\Omega
_{R}/|\nu |$. We clearly see that, for $|\nu |/\Omega _{R}\rightarrow 0$,
only one condensate survives, while in the opposite limit $\Omega _{R}/|\nu
|\rightarrow 0$, both condensates are essentially equal, with small
corrections of order $\Omega _{R}/|\nu |$.

The results presented in this paper are valid, provided the two-body
interaction is invariant under isospin rotations. Consider, for instance, a
small deviation from the $O(2)$ model, $g_{\phi }=g_{\psi }=g$, but $g_{\psi
\phi }=g+\Delta g$. Upon rotation to the $(\varphi _{1},\varphi _{2})$
basis, a term proportional to $\Delta g(\varphi _{1}^{\ast }\varphi
_{1}^{\ast }\varphi _{2}\varphi _{2})$ will be generated. Thus, even though
we have ignored this type of terms in the original model, they will be
generated in a more general two-body interaction case. Thus, for $\Delta
g\neq 0$, there is no isospin direction in which the $U(1)\otimes U(1)$
symmetry is recovered. This fact makes the study of quantum and thermal
fluctuations more involved. We hope to report on this issue shortly.

\section*{Acknowledgments}
The Brazilian agencies \emph{Conselho Nacional de Desenvolvimento 
Cient\'{\i}fico e Tecnol\'{o}gico (CNPq)} , \emph{Funda{\c{c}}{\~{a}}o de Amparo {\`{a}} 
Pesquisa do Estado do Rio de Janeiro (FAPERJ)} and \emph{ Coordena\c c\~ao de 
Aperfei\c coamento de Pessoal de N\'\i vel Superior (CAPES)} are acknowledged for
partial financial support. V.C.S. was financed by a doctoral fellowship by
CAPES. Z.G.A. was partially financed by a post-doctoral fellowship by CAPES. 
D.G.B. acknowledge support from  the Abdus Salam International Centre for Theoretical
Physics, ICTP, Trieste as a senior associated.

\appendix
\section{Rabi frequency and detuning}
\label{app}
Although  Rabi frequency and detuning are very well known concepts in 
atomic physics and Raman spectroscopy,  we would like to sketch in this appendix  
a brief summary relevant for  the  definition of our model.

The general context is the study of transition probabilities between hyperfine atomic states induced by an electromagnetic interaction. 
Just to keep matters simple,  consider, for instance, a two-level quantum system interacting with a {\em classical} electromagnetic field. The eigenstates of the free Hamiltonian are characterized by the two-dimensional orthogonal basis 
$\{|1\rangle,|2\rangle\}$, in such a way that the energies $E_1, E_2$ are the eingenvalues of the free Hamiltonian $H_0$, 
\begin{eqnarray}
H_0|1\rangle&=&E_1 |1\rangle, \\
H_0|2\rangle&=&E_2 |2\rangle
\end{eqnarray} 
and $\langle 1|1\rangle=\langle 2|2\rangle=1$, $\langle 1|2\rangle=\langle 2|1\rangle=0$.
 In this way, a general time dependent state can be written as
 \begin{equation}
 |\psi(t)\rangle= C_1(t) |1\rangle +  C_2(t) |2\rangle 
 \end{equation} 
with $|C_1|^2+|C_2|^2=1$.
Defining the spinor  $\psi(t)=\{C_1(t),C_2(t)\}$, the Shr\"odinger equation reads 
\begin{equation}
i\hbar \frac{d\psi}{dt}=\hat H\psi \ ,
\label{eq:Shodinger}
\end{equation}
with $\hat H=\hat H_0+\hat H_{\rm int}$
and 
\begin{equation}
\hat H_0=\left(
\begin{array}{cc}
E_1 & 0 \\
0 &  E_2 
\end{array}
\right).
\end{equation}
To built the interaction Hamiltonian we consider that the electromagnetic field  induces a dipole moment between the states 
$|1\rangle$ and $|2\rangle$ and the electric field couples with this dipole moment in such a way that 
\begin{equation}
\hat H_{\rm int}=\left(
\begin{array}{cc}
0 & \hbar \Omega_R \cos(\omega t +\alpha) \\
\hbar \Omega_R \cos(\omega t +\alpha) &  0
\end{array}
\right),
\end{equation}
where $\hbar \Omega_R$ is the dipole interaction energy, $\omega$ and $\alpha$ are the frequency and phase of the electromagnetic field respectively. The coupling $\Omega_R$ is usually called the Rabi frequency, while the frequency 
$\omega=(E_1-E_2)/\hbar + |\nu|$, where $|\nu|$ is the detuning of the frequency related with the resonance frequency  
$\omega_0=(E_1-E_2)/\hbar$.
If we consider that at the initial time $t=0$ the system is in the ground state $|1\rangle$, we can easily solve the equation~(\ref{eq:Shodinger}) with the initial conditions $C_1(0)=1,C_2(0)=0 $, finding 
\begin{equation}
|C_1(t)|^2=\left(\frac{|\nu|}{\Omega_{\rm eff}}\right)^2\left[1+ 
\left(\frac{\Omega_R}{|\nu|}\right)^2\cos^2\left(\frac{\Omega_{\rm eff} t}{2}\right)
\right],
\end{equation}
where ${\Omega_{\rm eff}}=\sqrt{\Omega_R^2+|\nu|^2}$ is called the effective Rabi frequency and $|C_2(t)|^2=1-|C_1(t)|^2$.

Thus, the dynamical behavior of the two-level system is driven  by two parameters, the Rabi frequency $\Omega_R$, which measures the coupling strength of the dipole with the electromagnetic field, and the detuning $|\nu|$, which measures the distance between the frequency of the applied field and the resonance frequency $\omega_0=(E_1-E_2)/\hbar$. Notice that the time dependency is completely given by the effective Rabi frequency $\Omega_{\rm eff}$, while the ratio $\Omega_R/|\nu|$ controls the amplitude of the probability density. 

Consider the system near the resonance (very small detuning) and in a weak coupling  regime (small Rabi frequency). Then, the usual \emph{rotating wave approximation} can be performed. It consists in writing the Hamiltonian in the interaction picture discarding rapidly fluctuating terms (terms that oscillates with $2\omega$). In this approximation, the Hamiltonian takes the simpler form 
\begin{equation}
\hat H=\left(
\begin{array}{cc}
|\nu| &  \Omega_R e^{i\alpha} \\
\Omega_R e^{-i\alpha}  & -|\nu|
\end{array}
\right)
\end{equation}
(where we have set $\hbar=1$).
This form of the one-particle Hamiltonian was used  in Ref.~\cite{DanWei} to describe effective spin-orbit interactions in two bosonic species systems. 

Equivalently, we can make another unitary transformation of the form
\begin{equation}
\hat H'= U^\dagger \hat H U \ ,
\end{equation}
with 
\begin{equation}
U=\frac{i}{\sqrt{2}}\left(
\begin{array}{cc}
e^{i\alpha} &  1 \\
1  &- e^{-i\alpha}
\end{array}
\right)\;
\end{equation}
and  $U^\dagger U=I$, obtaining
\begin{equation}
\hat H'=\left(
\begin{array}{cc}
\Omega_R &  |\nu| e^{-i\alpha} \\
|\nu| e^{i\alpha}  &- \Omega_R
\end{array}
\right)\;.
\end{equation}
This form of the one-body Hamiltonian was considered in Ref.~\cite{Garcia-March} to study spin-orbit couplings and it is the form  we have adopted to build our model. 

The model discussed in our paper is evidently more complex than the simple model described in this appendix, since it is composed by \emph{two interacting fields}.  While the condensates 
could be considered in some approximation as a two-level system, fluctuations out of the condensate strongly renormalized the bare parameters $\Omega_R$ and $\nu$. We showed that the effective Rabi frequency $\Omega_{\rm eff}$ is strongly renormalized by temperature, while the ratio $\Omega_R/|\nu|$ is unaffected by thermal fluctuations.


\end{document}